\documentclass[twocolumn,aps,prd,amsmath,amssymb,floatfix,amsmath,superscriptaddress]{revtex4-1}

\usepackage{graphicx}
\usepackage{dcolumn}
\usepackage{bm}
\usepackage{braket}
\usepackage{color,epsfig}
\usepackage{bm}
\usepackage{slashed}
\usepackage{hyperref}
\usepackage{subfigure}
\usepackage{feynmf}
\usepackage{calrsfs}

\newcommand{\bea}{\begin{eqnarray}}
\newcommand{\eea}{\end{eqnarray}}
\newcommand{\be}{\begin{equation}}
\newcommand{\ee}{\end{equation}}

\renewcommand\vec{\bm}

\begin{document}


\title{Corrections to the Gyromagnetic Factor\\
in
Very Special Relativity
}

\author{Benjamin Koch}
 \email{benjamin.koch@tuwien.ac.at}
\affiliation{Institut f\"ur Theoretische Physik and Atominstitut,
 Technische Universit\"at Wien,
 Wiedner Hauptstrasse 8--10,
 A-1040 Vienna, Austria}
\affiliation{Facultad de F\'isica, Pontificia Universidad Cat\'olica de Chile, Vicu\~{n}a Mackenna 4860, Santiago, Chile}
 
\author{Enrique Mu\~{n}oz}
\email{munozt@fis.puc.cl}
\affiliation{Facultad de F\'isica, Pontificia Universidad Cat\'olica de Chile, Vicu\~{n}a Mackenna 4860, Santiago, Chile}

\author{Alessandro Santoni}
\email{asantoni@uc.cl}
\affiliation{Facultad de F\'isica, Pontificia Universidad Cat\'olica de Chile, Vicu\~{n}a Mackenna 4860, Santiago, Chile}


\date{\today}

\begin{abstract}
We consider the corrections arising from the SIM(2) invariant realization of Very Special Relativity to the energy spectrum of a $C-$invariant Dirac Fermion in a static and homogeneous magnetic field $\vec B$. First, we analyze the case of $\vec B$ parallel to the spatial VSR preferred direction $\vec n$, finding that the expression for the energy spectrum stays the same, except for a mass shift arising from the VSR contribution. Then, we relax the parallelism condition, finding a new equation for the energy spectrum. We solve this equation perturbatively. With a Penning trap's experiment in mind, we derive the first order VSR corrections to the electron's $g-2$ factor. Finally, using the most accurate electron's $g$-factor measurements in Penning trap's experiments, we obtain an upper bound to the VSR electron mass parameter, and therefore also to the VSR electronic neutrino mass, of $~1 \, eV$. This result does not contradict the possibility for VSR to be the origin of neutrino masses. 
\end{abstract}

\maketitle

\section{Introduction}

Very Special Relativity (VSR) \cite{vsr1} is a theory where the flat spacetime symmetries are reduced to a subgroup of the Lorentz group plus the group of spacetime translations, which it is kept unchanged. While its classical consequences are, to our current knowledge, identical to the ones implied by special relativity, the non-classical consequences have already been analyzed in several physics' areas. 
The original idea that motivated Cohen and Glashow to formulate VSR was a new mechanism for the emergence of Neutrino's masses \cite{vsr2}.
Further applications of VSR that have been studied regard Supersymmetric extensions \cite{ss1,ss2}, the Gravitational sector \cite{grav1,grav2,grav3}, Quantum Electrodynamics \cite{qed1,qed2}, the Standard Model of Particle Physics~\cite{dunn2006,alfaro2}, and much more. \\
A minimal candidate for VSR is the $T(2)$ subgroup, that contains the generators $T_1 = K_1 + J_2$, and $T_2 = K_2 - J_1$, that combine boosts $\mathbf{K}$ and rotations $\mathbf{J}$. When $T(2)$ is enlarged to include $J_3$, the resulting subgroup is isomorphic to the 3-parameter group of Euclidean translations, $E(2)$. On the other hand, if $T(2)$ is enlarged to incorporate $K_3$, the resulting 3-parameter subgroup is isomorphic to the homotheties group $HOM(2)$. Finally, if $T(2)$ is equipped with both $K_3$ and $J_3$ generators, the resulting four parameter subgroup is isomorphic to the similitudes group $SIM(2)$. Each of these four VSR subgroups can be expanded to the full Lorentz group by the addition of discrete symmetries $P$, $T$, $CP$, or $CT$. In the rest of this work, we shall focus on the $SIM(2)$ subgroup of the Lorentz group, since it is the biggest proper Lorentz subgroup and it directly preserves CPT symmetry \cite{vsr1}, an important ingredient in all the known quantum field theories. Therefore, in the rest of the paper we will refer to this particular $SIM(2)$-realization of VSR just as VSR. 
\\
The main feature of the VSR theory is the introduction of a light-like preferred spacetime direction $n^\mu = (n^0, \vec n)$, which under $SIM(2)$ transformations changes only by a scaling factor
\begin{equation}
    n^\mu \underset{SIM(2)}{ \longrightarrow} e^{ \phi} n^\mu \, .
\end{equation}
Therefore, ratios of scalar products with $n^\mu$ both in the numerator and denominator are invariant under $SIM(2)$ transformations, but not under the full Lorentz group. 
The VSR idea, as it is implemented into the Dirac equation~(\ref{VSRdirac}), is a non-local violation of Lorentz symmetry and thus different from the more well known approach proposed by Kostelecky~\cite{kost,kost2}, where Lorentz violation is generally parametrized by spurion fields, which in this particular realization of VSR are not present.
\\
The focus of this paper will be to consider the effect of a static and homogeneous magnetic field $\vec B$ on the energy spectrum of a Dirac Fermion in the framework of VSR.
Furthermore, since transition energies between these energy levels are measured in experiments with Penning traps to calculate the gyromagnetic factor of electrons \cite{penning1,penning2,penning3}, we can estimate the effect of VSR corrections to the g-factor as well.
\\
The remaining of the paper is structured as follows: in Section \ref{sec2}, we analyze the case in which the VSR spatial vector $\vec n$ is parallel to the direction of the magnetic field $\vec B$, thus finding the corresponding exact energy eigenvalues. 
In section \ref{sec3}, we study the more general case, allowing an angular spacing, labelled by the angle $\theta$, between the direction of $\vec n$ and the magnetic field. 
In section \ref{sec4}, considering the experimental setup of an electron Penning trap, we study the possible consequences of VSR on the measured gyromagnetic moment of the electron, finding corrections even at zero order in the magnetic field. 
Finally, in section \ref{concl} we state our conclusions about the results and the possible future applications. In the appendices \ref{app1}, \ref{app2}, \ref{app3} and \ref{Borel} we show, respectively, the explicit calculations involved in the equation of motion for the upper spinor $\varphi(x^1)$, the derivation of the integrals $I_1(n,k)$, $I_2(n,k)$, the calculation of the perturbation matrix elements $V^{\bar n}_{\alpha, \alpha'}$, and some details on Borel regularization.

\section{Magnetic Field parallel to VSR direction} \label{sec2}

Let's start working in the VSR framework by considering the C-symmetric Dirac equation for a charged fermion~\cite{dunn2006}
\begin{eqnarray} 
\left(i\slashed{\partial} - m + i \frac{M^2}{2}\slashed{N} \right)\psi(x) = 0 \, ,
\label{VSRdirac}
\end{eqnarray}
where $N^\mu = n^\mu /n \cdot \partial $.
The VSR correction in this equation is
controlled by the parameter $M$ such that
for $M\rightarrow 0$ the full Lorentz symmetry is recovered.
Now, we include the external, constant and uniform magnetic field $\mathbf{B} = B\, \mathbf{\hat e}_3 $ by the usual minimal substitution $\partial_\mu \to \partial_\mu +i e A_\mu$, being $A_\mu$ the electromagnetic four-potential
\begin{eqnarray}
\left(i\slashed{\partial} - e\slashed{A} - m + i \frac{M^2}{2}\slashed{N} \right)\psi(x) = 0 \, ,
\label{eq_EOM}
\end{eqnarray}
such that the VSR operator becomes
\begin{equation}\label{eq_NVSR}
    N^{\mu} = \frac{n^{\mu}}{n\cdot\left(\partial + i e A  \right) }\,.
\end{equation}
\\
Due to the $n^\mu$-rescaling simmetry of Eq.~\eqref{VSRdirac}, we can always choose it in the form $n^\mu=(1, \mathbf{\hat n})$, with $|\mathbf{\hat n}|^2=1$. Here, in particular, we define the ``preferred'' VSR null vector as
\begin{eqnarray}
n^{\mu} = \left(1,0,0,1\right) \rightarrow n\cdot n=0 \,.
\label{eq_n}
\end{eqnarray}
In what follows, we choose the metric $g^{00}=+1$, $g^{ij}=-\delta_{ij}$ for $i=1,2,3$.
To represent the uniform magnetic field $\mathbf{B}$, we choose the gauge 
\begin{eqnarray}
A^{0} = A^{1} = A^{3} = 0,\, A^{2}(x^{1}) = B x^{1} \, ,
\label{eq_gauge}
\end{eqnarray}
such that translational symmetry along $ \mathbf{\hat e}_2$ and $\mathbf{\hat e}_3$ is preserved
\begin{eqnarray}
\left[A^{\mu}(x),p_3 \right]_{-} = \left[A^{\mu}(x),p_2 \right]_{-} = 0 \,,
\label{eq_comm}
\end{eqnarray}
where the components of the momentum operator are defined as $p_j = -p^j = i\partial_{j}$.
By taking this into account, the 2 and 3-components of the momentum can be susbstituted by their eigenvalues, i.e.
$p_3 \rightarrow k_3$ and $p_2 \rightarrow k_2$. Moreover, we have that an eigenstate of \eqref{eq_EOM} is of the form
\begin{equation}
\psi(x) = e^{-i E t} e^{i\left( k^3 x^3 + k^2 x^2 \right)}\left(\begin{array}{c}\varphi(x^1)\\\chi(x^1)\end{array}\right) \, .
\label{eq_eig1}
\end{equation}
For the gauge \eqref{eq_gauge} we also have that $n\cdot A =0$ automatically, and hence the VSR operator \eqref{eq_NVSR} for an eigenstate of the form \eqref{eq_eig1} simplifies to
\begin{eqnarray}
i\frac{M^2}{2}\slashed{N} \rightarrow -\frac{M^2}{2}\frac{\gamma^0 - \gamma^3}{E - k^3}\, ,
\label{eq_VSR2}
\end{eqnarray}
where $\gamma^{\mu}$ are the Dirac matrices, for which we choose the standard representation
\begin{eqnarray}
\gamma^0 = \left( \begin{array}{cc}\mathbf{1} & 0\\0 & -\mathbf{1} \end{array}\right),\,\,\gamma^i = \left( \begin{array}{cc}0 & \sigma^i\\-\sigma^i & 0 \end{array}\right) \,,
\label{eq_gamma}
\end{eqnarray}
where $\mathbf{1}$ is a two by two identity matrix and $\vec \sigma$ the usual Pauli's matrices.
\subsection{Equations of motion for $\varphi(x^1)$}

After inserting Eq.~\eqref{eq_eig1} and Eq.~\eqref{eq_VSR2} into Eq.~\eqref{eq_EOM}, and dividing the bi-spinor $\psi$ in its upper and lower components as
\begin{equation}
    \psi(x^1) = \left(\begin{array}{c}\varphi(x^1)\\\chi(x^1) \end{array}\right) \, ,
\end{equation}
we obtain the following equation
\begin{widetext}
\begin{eqnarray}
\left[\begin{array}{cc} \left(E - m - \frac{M^2}{2\left(  E - k^3\right)}  \right)\mathbf{1} & e B x^1 \sigma^2 +
\frac{M^2}{2\left(  E - k^3\right)}\sigma^3 - \sigma^i p^i\\-e B x^1 \sigma^2 -
\frac{M^2}{2\left(  E - k^3\right)}\sigma^3 + \sigma^i p^i & -\left(E + m - \frac{M^2}{2\left(  E - k^3\right)}  \right)\mathbf{1}
\end{array} \right] \left(\begin{array}{c}\varphi(x^1)\\\chi(x^1) \end{array}\right) = 0 \, ,
\label{eq_Dirac1}
\end{eqnarray}
\end{widetext}
From the system \eqref{eq_Dirac1}, we solve for the lower spinor $\chi$ in terms of the upper one $\varphi$, to obtain the expression
\begin{eqnarray}
\chi(x^1) = -\frac{eB x^1\sigma^2 + \frac{M^2}{2(E - k^3)}\sigma^3 -\sigma^i p^i}{E + m - \frac{M^2}{2(E - k^3)}}\varphi(x^1) \, .
\label{eq_chi}
\end{eqnarray}
Inserting Eq.~\eqref{eq_chi} back into the first system's equation of \eqref{eq_Dirac1}, we obtain an expression for the upper spinor $\varphi$
\begin{eqnarray} \label{eq_phi}
&&\left[\left( E - \frac{M^2}{2\left( E - k^3  \right)} \right)^2 - m^2\right.\\
&&\left.- \left(eB x^1 \sigma^2 + \frac{M^2}{2(E - k^3)}\sigma^3 - \sigma^i p^i  \right)^2  \right]
\varphi(x^1) = 0 \, .\nonumber
\end{eqnarray}
By applying the standard properties of the SU(2) algebra, we can calculate the square of the differential operator in Eq.~\eqref{eq_phi}, to obtain
\begin{eqnarray}
&&\left[\left( E - \frac{M^2}{2\left( E - k^3  \right)} \right)^2 - m^2- \left( \frac{M^2}{2(E-k^3)}-k^3 \right)^2\right.\nonumber\\
&&\left. -\left(e B x^1 - k^2  \right)^2  +  e B\sigma^3 - p_1^2 \right] \varphi(x^1) = 0 \, .
\label{eq_phi2}
\end{eqnarray}
Clearly, the equation above is diagonal in the two components of the spinor $\varphi(x^1)$
\begin{eqnarray}
\varphi(x^1) = \left(\begin{array}{c} f^{1}(x^1)\\f^{2}(x^1) \end{array} \right) \, .
\end{eqnarray}
Therefore, the eigenvalue problem will have two independent solutions, that we define by 
\begin{equation}
    \vec f_{+1} (x^1) \equiv \left(\begin{array}{c} f_{+1}(x^1)\\0 \end{array} \right) \, , \,\,\,
    \vec f_{-1} (x^1) \equiv  \left(\begin{array}{c} 0\\f_{-1}(x^1) \end{array} \right)\,.
\end{equation}
The $f_\alpha (x^1)$ introduced above, with $\alpha=\pm 1$ representing the two eigenvalues of $\sigma^3$, are given by the solution to the two independent differential equations
\begin{eqnarray}
&&\left[ -\partial_{1}^2 + \left(e B x^1 - k^2  \right)^2- \alpha e B - e B a(k_3,E) \right]f_{\alpha}(x^1)=0\, ,\nonumber\\
\label{eq_fa1}
\end{eqnarray}
where we defined the coefficient
\begin{eqnarray}
a(k^3,E) = \frac{1}{e B} &&\left[  \left(E - \frac{M^2}{2(E-k^3)}  \right)^2 - m^2\right.\nonumber\\
&&\left.- \left(k^3 - \frac{M^2}{2(E-k^3)}  \right)^2   \right] \, .
\label{eq_a}
\end{eqnarray}
It is convenient to define the dimensionless coordinate
\begin{eqnarray}
\xi = \sqrt{eB}\left(x^1 - \frac{k^2}{e B} \right) \,,
\end{eqnarray}
such that Eq.~\eqref{eq_fa1} becomes (for $\alpha = \pm 1$)
\begin{eqnarray}
\left[-\frac{d^2}{d\xi^2} + \xi^2 - \alpha \right]f_{\alpha}(\xi) = a(k^3,E) f_{\alpha}(\xi)\,.
\label{eq_falpha}
\end{eqnarray}
The only $L^2$-normalizable solutions of Eq.~\eqref{eq_falpha} are the functions
\begin{eqnarray}\label{def_falpha}
f_{n,\alpha}(\xi) = C e^{-\xi^2/2} H_{n}(\xi)\,,
\end{eqnarray}
where $H_n(\xi)$ are the Hermite polynomials of order $n\in\mathbb{N}_0$, provided the following quantization condition is satisfied
\begin{eqnarray}
a(k^3,E) + \alpha = 2 n + 1,\,n\in\mathbb{N}_0,\,\alpha=\pm 1\,,
\label{eq_aq}
\end{eqnarray}
while, the normalization coefficient $C$ in Eq.~\eqref{def_falpha} is obtained from the orthonormality condition for the Hermite polynomials
\begin{equation}
    \int _{-\infty}^{+\infty} d\xi \; e^{-\xi^2} H_n(\xi) H_m(\xi) = 2^n n! \sqrt{\pi} \delta_{nm}\,.
    \label{eq_normHerm}
\end{equation}

\subsection{Energy Spectrum}

Following Eq.~\eqref{eq_aq}, the energy spectrum $E \equiv E_{\pm}\left(k^3,n,\alpha \right)$ is defined by the roots of the algebraic equation
\begin{eqnarray}
\left(E - \frac{M^2}{2(E - k^3)}\right)^2 &-& \left(k^3-\frac{M^2}{2(E-k^3)}\right)^2 = \nonumber\\
&=& e B\left(2 n + 1 - \alpha\right) + m^2 \,.
\end{eqnarray}
This equation can be solved explicitly, to obtain the exact energy eigenvalues
\begin{eqnarray}
E^{(0)}_{\pm}(k^3, n,\alpha) &=& \pm \sqrt{e B (2n + 1 - \alpha) + (k^3)^2 + m^2_f }\, ,\nonumber\\
\label{eq_VSRspectrum}
\end{eqnarray}
where
\begin{equation}
m^2_f= m^2 +M^2.
 \label{mf}
\end{equation}
One confirms that when the full Lorentz symmetry is restored  in limit $M^2\rightarrow0$, 
the spectrum in Eq.~\eqref{eq_VSRspectrum} reduces to the well known ``unperturbed" solutions
\begin{eqnarray} \label{E0}
\left.E^{(0)}_{\pm} (k^3, n,\alpha)\right|_{M=0} = \pm  E^{(u)}(k^3, n,\alpha)  \,,
\end{eqnarray}
with 
\begin{eqnarray}
    E^{(u)}(k^3, n,\alpha) = \sqrt{e B (2n + 1 - \alpha) + (k^3)^2 + m^2}\,.\nonumber \\
\end{eqnarray}
Moreover, it is also clear from Eq.~\eqref{eq_VSRspectrum} that for this configuration, where the field $\mathbf{B}$ is parallel to the direction of $  {\mathbf{ \hat n}}$, the sole effect of the VSR term in the single-particle spectrum is to shift the particle's mass $m \rightarrow m_f$.\\ 
Following the analysis in the previous section, we remark that, apart from the ground state energy with $n=0$ and $\alpha=+1$, each energy eigenvalue is degenerate since we can obtain it with the two combinations $(n\, ,\alpha= -1)$ and $(n+1 \,, \alpha=+1)$.\\
Therefore, introducing the system's eigenstates $\ket{\vec f^{(0)}} $ such that $\vec f^{(0)} (\xi) = \langle\xi|\vec{f} ^{(0)}\rangle$, our eigenvector's basis will look like
\begin{eqnarray}
    \bigg \{
    \ket{\vec{f}^{(0)}, 0,+ 1} ,&&  \left [\ket{\vec{f}^{(0)},0,- 1}, \ket{\vec{f}^{(0)},1,+ 1} \right ]\, , \,...\, , \\
    && \left [\ket{\vec{f}^{(0)},n,- 1}, \ket{\vec{f}^{(0)},n+1,+ 1} \right ] ,\,...
    \bigg \} \nonumber \, , 
\end{eqnarray}
with the square brackets highlighting the 2-dimensional degenerate eigenspaces. To simplify the notation in the rest of the calculations and to label each degenerate eigenspace, we re-order the eigenstates by $\ket{\vec f^{(0)} , n , \alpha} \to \ket{\vec f^{(0)} , \bar n, \alpha} $, where
\begin{equation}
    \left\{\begin{array}{l} 
    \bar n = n \;\;\;\;\;\;\;\;\; for \; \alpha =+1\\ 
    \bar n= n+1 \;\;\; for \; \alpha =-1 \; . 
    \end{array} \right.  
\end{equation}
The eigenvector's basis then becomes
\begin{eqnarray}
    \bigg \{
    \ket{\vec{f}^{(0)}, \bar 0,+1  } ,&&  \left [\ket{\vec{f}^{(0)},\bar 1,-1 }, \ket{\vec{f}^{(0)},\bar 1,+ 1} \right ]\, , \,...\, , \\
    && \left [\ket{\vec{f}^{(0)}, \bar n,-1 }, \ket{\vec{f}^{(0)}, \bar n,+1 } \right ] ,\,...
    \bigg \} \nonumber \, , 
\end{eqnarray}
where each degenerate eigenspace is spanned by the two orthogonal eigenvectors
\begin{eqnarray} \label{eq_unpertspinor}
\langle\xi|\vec{f}^{(0)},\bar n,+1\rangle &=& \frac{1}{\pi^{1/4}2^{\frac{\bar n}{2}}\sqrt{\bar n!}}\left( \begin{array}{c}e^{-\frac{\xi^2}{2}}H_{\bar n}(\xi)\\0 \end{array}\right)\,, \\
\langle\xi|\vec{f}^{(0)}, \bar n,-1\rangle &=& \frac{1}{\pi^{1/4}2^{\frac{\bar n-1}{2}}\sqrt{(\bar n-1)!}}\left( \begin{array}{c}0\\e^{-\frac{\xi^2}{2}}H_{\bar n-1}(\xi) \end{array}\right) \,, \nonumber
\end{eqnarray}
while the ground state $\bar n=n=0$ is defined by
\begin{equation}\label{ground0}
\langle\xi|\vec{f}^{(0)},\bar 0,+1\rangle = \frac{1}{\pi^{1/4}}\left( \begin{array}{c}e^{-\frac{\xi^2}{2}}H_{0}(\xi)\\0 \end{array}\right)\,,
\end{equation}
and its energy is not degenerate.

\section{Magnetic Field not parallel to VSR direction} \label{sec3}

Let us now consider a magnetic field $\mathbf{B}$ oriented with an angle $\theta \in [0,\pi]$ with respect to the VSR unit vector $  {\mathbf{\hat n}}$, i.e. $\mathbf{B}\cdot  {\mathbf{\hat n}} = B\cos\theta$. Therefore, without loss of generality, we choose the coordinate system such that $\mathbf{B} = B\, \mathbf{\hat e_3}$, and
\begin{eqnarray}
n^{\mu} = (1,  {\mathbf{\hat n}}) = (1,\sin\theta,0,\cos\theta) \, .
\end{eqnarray}
We still choose the same gauge as in the previous case, $A^{\mu} = (0,0,B x^1,0)$, so that the translational invariance along the $x^2$ and $x^3$ directions allows us to choose the same separation of variables as in Eq.~\eqref{eq_eig1}, while preserving the light-cone condition $n \cdot A=0$. \\
In this case, however, the VSR term in the equation of motion reduces to the form
\begin{eqnarray}
i \frac{M^2}{2}\slashed{N} \rightarrow -\frac{M^2}{2}\frac{\gamma^0 E - \gamma^1\sin\theta - \gamma^3\cos\theta }{E - k^3\cos\theta -  p^1 \sin\theta} \, .
\label{eq_VSR2-2}
\end{eqnarray}

\subsection{Equations of motion for $\varphi(x^1)$}

Substituting Eq.~\eqref{eq_VSR2-2} into Eq.~\eqref{eq_EOM}, we obtain for this general case the expression
\begin{eqnarray}
&&\left[\gamma^0 E + \gamma^1 p_1 - \gamma^2\left( k^2 - e B x^1 \right) - \gamma^3 k^3 - m\right.\\
&&\left.-\frac{M^2}{2}\frac{\gamma^0 E - \gamma^1\sin\theta - \gamma^3\cos\theta }{E - k^3\cos\theta -  p^1 \sin\theta}
\right]\left(\begin{array}{c}\varphi(x^1)\\\chi(x^1) \end{array}\right) = 0 \,.\nonumber
\label{eq_EOM2}
\end{eqnarray}

From this linear system, the lower spinor $\chi(x^1)$ can be solved in terms of the upper component $\varphi(x^1)$, as done in the case $\mathbf{B} \parallel \mathbf{\hat n}$, and then replaced again to obtain the following single equation for the upper spinor $\varphi(x^1)$
\begin{widetext}
\begin{eqnarray}
&&\left[\left( E - \frac{M^2/2}{E - k^3\cos\theta -  p^1 \sin\theta} \right)^2 - m^2\right.\nonumber\\
&&\left.- \left( \left(p^1 - \frac{M^2 \sin\theta/2}{E - k^3\cos\theta -  p^1 \sin\theta}\right)\, \sigma^1+ \left(k^2 - e B x^1 \right)\sigma^2 + \left(k^3 - \frac{M^2 \cos\theta/2}{E - k^3\cos\theta -  p^1 \sin\theta} \right)\sigma^3 \right)^2
\right]\varphi(x^1) = 0 \, .
\label{eq_phi3}
\end{eqnarray}
\end{widetext}
By expanding the squares, we observe that, due to the anti-commutation relations satisfied by the Pauli matrices, among the terms with mixed $\sigma$'s only the ones that involve operators acting on $x^1 \varphi(x^1)$ can have a chance to generate a surviving part after summing up the sigma products with interchanged indices.\\
For example, for terms involving $ p^1$, we can exploit the fact that, for any function $g(x^{1})$
\begin{equation} \label{anticom2}
\left[\sigma^{1} p^1,g(x^1)\sigma^2 \right]_{+} \varphi (x^1) = -\sigma^{3}\partial_{1}g(x^1) \varphi (x^1)\, .
\end{equation}
Applying $\sigma$'s anticommutation rules, the relation Eq.~\eqref{anticom2} and keeping in mind the representation Eq.~\eqref{intrepr} for the inverse operator, Eq.~\eqref{eq_phi3} is reduced to the expression (see appendix \ref{app1} for an explicit calculation)
\begin{widetext}
\begin{eqnarray} \label{eq_phi4}
&&\left[
E^2 - (k^3)^2 - m^2  -M^2 - ( p^1)^2 + e B  \sigma^3 - (k^2 -e B x^1)^2  \right.\nonumber\\
&&\left. - \frac{M^2}{2} e B \frac{\sin^2 \theta}{(E - k^3\cos\theta -  p^1 \sin\theta)^2}  \sigma^3 + \frac{M^2}{2} e B \frac{\sin\theta \cos\theta}{(E - k^3\cos\theta -  p^1 \sin\theta)^2}  \sigma^1
\right]\varphi(x^1) = 0 \, .
\end{eqnarray}
\end{widetext}
Now, let us define:
\begin{eqnarray} \label{aEdef}
&&a(k^3, E)= \frac{1}{eB}\left(E^2 - (k^3)^2 - m^2 - M^2
\right) \, , \nonumber \\
&&\tilde E = E -k^3 \cos\theta \, ,
\end{eqnarray}
along with the change of variables
\begin{eqnarray}
\xi = \sqrt{e B}\left(x^1 - k^2/(eB) \right) \, .
\end{eqnarray}
Therefore, dividing Eq.~\eqref{eq_phi4} by $e B$ and using the above definitions, we obtain
\begin{eqnarray}
&& \left[-\partial_{\xi}^2 + \xi^2 - \sigma ^3  -a (k^3,E) \right.\\
&& +\left. \frac{M^2 \sin\theta /2}{(\tilde E - \sqrt{e B} p^\xi \sin\theta)^2} (\sin\theta  \sigma^3 - \cos\theta  \sigma^1) \right] \varphi(x^1) = 0 \, , \nonumber
\end{eqnarray}
which, defining the operator $P_\xi \equiv \tilde E + i \sqrt{e B}  \sin\theta \partial_\xi $, can be expressed as an eigenvalue equation
\begin{eqnarray}\label{eq_f1}
    && \left[ -\partial_{\xi}^2 + \xi^2 - \sigma ^3  + \frac{M^2 \sin \theta }{ 2  P^2_\xi} (\sin\theta\,  \sigma^3 \right. \nonumber  \\
    &&  \left. - \cos\theta \,  \sigma^1  ) \right ] \left(\begin{array}{c}f^{1}(\xi)\\f^{2}(\xi) \end{array}\right)=
    a \left(\begin{array}{c}f^{1}(\xi)\\f^{2}(\xi) \end{array}\right) \, .
\end{eqnarray}

\subsection{VSR Perturbative Scheme}

Thinking of VSR as a correction to special relativity, we can consider $M$ as a small parameter compared to the other system's energy scales, like $m$ or $\sqrt{eB}$, so that we identify an unperturbed operator $H_0$ and a perturbation operator $V$ 
\begin{eqnarray}
&& H_0 = -\partial_{\xi}^2 + \xi^2 - \sigma ^3  \nonumber \\
&& V = \frac{\sin\theta}{   P^2_\xi} (\sin\theta  \sigma^3 - \cos\theta  \sigma^1) \, .
\end{eqnarray}
Thus, in terms of the system's eigenstates $\ket {\vec f}$, we can write Eq.~\eqref{eq_f1} as
\begin{equation}
    (H_0 + \frac{M^2}{2} V ) \ket{\vec f}= a \ket{\vec f}\, .
\end{equation}
We can then approach the problem in a perturbative scheme. Let's define $\lambda := M^2/2$ and expand in a power series $\ket {\vec f}$ and $a$ so that
\begin{eqnarray}
 a &=& a^{(0)} + \lambda a^{(1)} + O(\lambda^2) \, ,\nonumber \\
\ket {\vec f} &=&  \ket{\vec f^{(0)}} + \lambda \ket{\vec f^{(1)}} + O(\lambda^2)  \, .
\end{eqnarray}
Therefore, we are considering a perturbation to the solution already obtained in the previous section 
\begin{equation}
    H_0 \ket{\vec f^{(0)},\bar n,\alpha} = a^{(0)}_{\bar n,\alpha} \ket{\vec f^{(0)}, \bar n,\alpha} \, ,
\end{equation}
with $\braket{\xi | \vec f ^{(0)} , \bar n , \alpha}$ defined by Eqs.~\eqref{eq_unpertspinor}, \eqref{ground0}, and $a^{(0)}_{n,\alpha} = 2n +1-\alpha$, or in the $\bar n -$basis $a^{(0)}_{\bar n,\alpha} =2 \bar n$.
At first order in the perturbation ${V}$, we find
\begin{equation}
    (H_0 -a^{(0)}_{\bar n,\alpha } \mathbf{1}) \ket{\vec f^{(1)} , \bar n ,\alpha} = (a^{(1)}_{\bar n, \alpha} \mathbf{1} - V) \ket{\vec f^{(0)} ,\bar n,\alpha}\,
    \label{eq_pert1}
\end{equation}
\subsubsection{Perturbative correction to the state $\bar n=0$}
For the case $\bar n=0$, we necessarily have $\alpha=+1$ and no degenerate eigenvectors, corresponding then to ordinary perturbation theory. Multiplying \eqref{eq_pert1} by $\bra{f^{(0)}, \bar 0,+1 }$, we see that
\begin{eqnarray}
    a^{(1)}_{\bar 0,+1} &=& \bra{\vec f^{(0)} ,\bar 0 ,+1}{V}\ket{ \vec f^{(0)} , \bar 0 ,+1}
\end{eqnarray}
Inserting two identities through the completeness relation $1= \int d\xi \ket{\xi}\bra{\xi}$ and using \eqref{ground0} we obtain
\begin{eqnarray}\label{a1n0.1}
a^{(1)} _{\bar 0,+1} = &&  \frac{1}{\sqrt{\pi} } \int_{-\infty}^{\infty} d\xi \; e^{-\xi^2/2} H_0(\xi) \frac{\sin^2\theta}{P^2_\xi} (e^{-\xi^2/2} H_0(\xi))  \nonumber\\
= && - \frac{\sin^2\theta}{ \sqrt{\pi} \tilde E^2 } \frac{d}{d A }\int_{-\infty}^{\infty} d\xi \; e^{-\xi^2/2} H_0(\xi) \times \nonumber\\
&& \times \frac{1}{A+i \eta\sin\theta \partial_\xi} (e^{-\xi^2/2} H_0(\xi)) \bigg |_{A=1} \, ,
\end{eqnarray}
where we have defined the dimensionless quantity
\begin{equation}
    \eta=\sqrt{e B}/{\tilde E}.
\end{equation}
For calculation purposes, we introduce a Schwinger-type integral representation for the inverse operator in Eq.~\eqref{a1n0.1} (valid for $A>0$)
\begin{eqnarray}
\frac{1}{A+i \eta\sin\theta \partial_\xi} = \int_{0}^{\infty}dt\,e^{-t\left(A+i \eta\sin\theta \partial_\xi\right)}.
\label{intrepr}
\end{eqnarray}
With the integral form in Eq.~\eqref{intrepr}, we get
\begin{eqnarray}
a^{(1)} _{\bar 0,+1} =&& - \frac{ \sin^2\theta}{\sqrt{\pi} \tilde E^2 } \frac{d}{d A}\int_{0}^\infty dt \int_{-\infty}^{\infty} d\xi \; e^{-\xi^2/2} H_0(\xi) \times \nonumber \\
&& \times  e^{- A t (1+i \eta\sin\theta \partial_\xi)} (e^{-\xi^2/2} H_0(\xi)) \bigg |_{A=1} =  \nonumber\\
=&& - \frac{\sin^2\theta}{\sqrt{\pi} \tilde E^2 } \frac{d}{dA}\int_{0}^\infty e^{-At} dt \int_{-\infty}^{\infty} d\xi \; e^{-\xi^2/2} H_0(\xi) \times \nonumber \\
&& \times  e^{-i t \eta\sin\theta \partial_\xi} (e^{-\xi^2/2} H_0(\xi)) \bigg |_{A=1} \,,
\end{eqnarray}
where clearly we take the limit $A\rightarrow 1$ at the end. 

Expanding the exponential operator
\begin{eqnarray}\label{a1n0.2}
a^{(1)} _{\bar 0,+1} =&& - \frac{\sin^2\theta}{ \sqrt{\pi} \tilde E^2 }  \sum_{k=0}^\infty \frac{(-i \eta\sin\theta )^k}{k!} \frac{d}{dA} \int_{0}^\infty e^{-At} t^k dt \times \nonumber  \\
&& \times  \int_{-\infty}^{\infty} d\xi \; e^{-\xi^2/2} H_0(\xi) \partial_\xi^k (e^{-\xi^2/2} H_0(\xi)) \bigg |_{A=1}  \nonumber\\
=&&  \frac{ \sin^2\theta}{ \sqrt{\pi} \tilde E^2 }  \sum_{k=0}^\infty (k+1) (i \eta\sin\theta )^k  I_1(0,k) \, ,
\end{eqnarray}
where we defined the integral
\begin{equation}\label{I1nk}
   I_1(\bar n,k)=  \int_{-\infty}^{\infty} d\xi \; e^{-\xi^2/2} H_{\bar n}(\xi) \partial_\xi^k (e^{-\xi^2/2} H_{\bar n}(\xi))\,.
\end{equation}
As shown in Appendix \ref{app2}, the parity properties of the integrand imply that it vanishes for odd values of k.\\
Moreover, as 
shown in detail in Appendix \ref{app2}, the analytical expression for the $\bar{n} = 0$ case is
\begin{eqnarray}
I_1(0,2k) = \sqrt{\pi} \, 2^{-2k}(-1)^k \frac{\Gamma(2k+1)}{\Gamma(k+1)}\,,
\end{eqnarray}
while for $\bar n>0$ we have
\begin{eqnarray}
I_1(\bar n,2k)=\sqrt{\pi} (-1)^k \bar n! \, 2^{\bar n-2k}\frac{\Gamma(2k+1)}{\Gamma(k+1)} F(-k,-\bar n;1;2)\,,\nonumber\\
\end{eqnarray}
where $F(a,b;c;z)$ is the Hypergeometric function.
Therefore, we can re-write Eq.~\eqref{a1n0.2} as
\begin{eqnarray} \label{a1n0sumk}
    a^{(1)} _{\bar 0,+1} &=& \frac{ \sin^2\theta}{ \sqrt{\pi} \tilde E^2 }  \sum_{k=0}^\infty (2k+1)(-1)^k \left(\eta\sin\theta \right)^{2k}  I_1(0,2k)\nonumber\\
    &=& \frac{ \sin^2\theta}{ \tilde E^2 } \sum_{k=0}^\infty \left(\frac{\eta\sin\theta}{2}\right)^{2k}\frac{\Gamma(2k + 2)}{\Gamma(k+1)},
\end{eqnarray}
which is a completely real expression as one would expect. The $k-$sum in Eq.~\eqref{a1n0sumk} is not convergent in the standard sense. However, it can be regularized, for example, by the Borel prescription, to obtain a closed form in terms of the incomplete Gamma function $\Gamma(-1/2,z)$ as follows (see Appendix \ref{Borel} for details)
\begin{eqnarray}\label{aln0borel}
a^{(1)} _{\bar 0,+1}= \frac{\sin^2\theta}{\tilde{E}}\frac{e^{-\frac{1}{\eta^2\sin^2\theta}}}{\left(- \eta^2\sin^2\theta\right)^{3/2}}  \,\Gamma \left( -\frac{1}{2},-\frac{1}{\eta^2\sin^2\theta}\right)\,.
\end{eqnarray}
At the lowest orders for $\eta\ll1$, both the power series Eq.\eqref{a1n0sumk} and the regularized Borel Eq.\eqref{aln0borel} reduces to,
\begin{eqnarray}
a^{(1)} _{\bar 0,+1}\simeq\frac{ \sin^2\theta}{ \tilde E^2 }\left[1 + \frac{3}{2}\eta^2\sin^2\theta + O(\eta^4)\right]\,.
\end{eqnarray}
We remark that, at this order $\eta^2$, the result is unique regardless of the regularization prescription. Then, to go further in our analysis, we assume to be in a situation where the magnetic field is small respect to other energy scales $\eta = \frac{\sqrt{eB}}{\tilde E}  \ll1$.\\
In this weak field approximation, we have that at first order in $\lambda$, going back to the $n- $notation (that does not make any difference for the ground state)
\begin{equation}\label{eqa10p}
    a_{0,+1} = a^{(0)} _{0,+1} + \lambda a^{(1)} _{0,+1} \approx \lambda \frac{\sin^2\theta}{\tilde E^2}\left[1 + \frac{3}{2}\eta^2\sin^2\theta  \right]\,.
\end{equation}

\subsubsection{Perturbative correction to the states $\bar n>0$}
For $\bar n>0$, instead, we notice that Eq.~\eqref{eq_pert1} corresponds to degenerate perturbation theory within the subspace spanned by the two degenerate spinors defined by Eq.~\eqref{eq_unpertspinor}, i.e. we have
\begin{eqnarray}
    \ket{\vec f^{(1)} ,\bar n ,\alpha} = \sum_{\alpha' = \pm} C^{\bar n}_{\alpha,\alpha'}\ket{\vec f^{(0)} ,\bar n ,\alpha'}.
    \label{eq_funcexp}
\end{eqnarray}
Therefore, projecting Eq.~\eqref{eq_pert1} over each of the spinors $\bra{\vec f^{(0)}, \bar n, \alpha'}$, and applying the zero-order property $\bra{\vec f^{(0)}, \bar n,\alpha}(H^0-a^{(0)}_{ \bar n,\alpha}     \mathbf{1}) = 0$, we obtain the linear eigenvalue system
\begin{equation}
    \left[\begin{array}{cc} V^{\bar n}_{+1,+1} & V^{\bar n}_{+1,-1}\\V^{\bar n}_{-1,+1} & V^{\bar n}_{-1,-1}\end{array}  \right]\left(\begin{array}{c}C^{\bar n}_{\alpha,+1}\\C^{\bar n}_{\alpha,-1}\end{array} \right) = a^{(1)} _{\bar n ,\alpha} \left(\begin{array}{c}C^{\bar n}_{\alpha,+1}\\C^{\bar n}_{\alpha,-1}\end{array} \right),
    \label{eq_Vsyst}
\end{equation}
where we defined the matrix elements of the perturbation within the subspace of unperturbed degenerate states
\begin{eqnarray}
    V^{\bar n}_{\alpha,\alpha'} = \bra{\vec f^{(0)} ,\bar n ,\alpha}  {V}\ket{\vec f^{(0)} ,\bar n ,\alpha'}.
\end{eqnarray}
Up to second order in $\eta$, the matrix is explicitly given by
\begin{equation}
    V^{\bar n} = \frac{\sin^2\theta }{\tilde E^2} \left[\begin{array}{cc}  1+\frac{3(1+2\bar n)}{2} \eta^2\sin^2\theta & i \eta \sqrt{\frac{\bar n}{2}} \cos\theta  \\ -i \eta \sqrt{\frac{\bar n}{2}} \cos\theta  & -(1+\frac{3(1+2\bar n)}{2} \eta^2\sin^2\theta) \end{array}  \right] \,.
\end{equation}
The calculation of the $ V^{\bar n}_{ \alpha,\alpha'}$, that follows from a similar procedure as in the previous $\bar n=0$ case, is shown in appendix \ref{app3}.\\
Therefore, the first-order correction $a^{(1)}_{\bar n,\alpha}$, defined up to order $\eta^2$, is obtained from the two eigenvalues of the linear system Eq.~\eqref{eq_Vsyst}, i.e. from the characteristic equation $\det(   V -      \mathbf{1} \, a_{\bar n}^{(1)}) = 0$, which reads
\begin{eqnarray}
   &&\left( \frac{a^{(1)}_{\bar n,\alpha} \tilde E^2 }{\sin^2\theta } +1+\frac{3(1+2\bar n)}{2} \eta^2\sin^2\theta\right) \left( \frac{a^{(1)}_{\bar n,\alpha} \tilde E^2 }{\sin^2\theta } -1\right.\nonumber\\
   &&\left.\;\;\;\;\;\;\;\;\;\;\;\;-\frac{3(1+2\bar n)}{2} \eta^2\sin^2\theta \right)  -\frac{\bar n}{2}\eta^2 \cos^2\theta = 0\,.
\end{eqnarray}
Solving for the two eigenvalues $a_{n,\alpha}^{(1)}$ up to order $\eta^2$, we obtain
\begin{eqnarray}
    a^{(1)}_{\bar n,\pm1} &=& \pm \frac{\sin^2\theta}{\tilde E^2} \sqrt{\left (1+\frac{3(1+2\bar n)}{2} \eta^2\sin^2\theta \right )^2 + \frac{\bar n}{2} \eta^2  \cos^2\theta }  \nonumber \\
    &\approx& \pm \frac{\sin^2\theta}{\tilde E^2} \left ( 1+ \eta^2\sin^2\theta ( \frac{3}{2} +3 \bar n +\frac{\bar n}{4} \cot^2\theta ) \right )\,.
\end{eqnarray}
In this perturbative scheme, we can identify the correction corresponding to each unperturbed eigenvector. Consider, for example, the positive eigenvalue $a^{(1)}_{\bar n,+1}$ of $V^{\bar n}$, for which we have the equation
\begin{eqnarray}
   \left[\begin{array}{cc}  (1+\frac{3(1+2\bar n)}{2} \eta^2\sin^2\theta) & i \eta \sqrt{\frac{\bar n}{2}} \cos\theta  \\ -i \eta \sqrt{\frac{\bar n}{2}} \cos\theta  & -(1+\frac{3(1+2\bar n)}{2} \eta^2\sin^2\theta) \end{array}  \right]  \left(\begin{array}{c}C^{\bar n}_{+1}\\C^{\bar n}_{-1}\end{array} \right)\nonumber\\
    = \left ( 1+ \eta^2\sin^2\theta ( \frac32 +3 \bar n +\frac{\bar n}{4} \cot^2\theta ) \right ) \left(\begin{array}{c}C^{\bar n}_{+1}\\C^{\bar n}_{-1}\end{array} \right). \nonumber\\
\end{eqnarray}
Here, applying the $L^2$-normalization $|C_{+1}^{\bar n}|^2+|C_{-1}^{\bar n}|^2=1$ and choosing the arbitrary phase such that $C_{+1}^{\bar n}$ is real, we obtain
\begin{eqnarray}
    &&C_{+1}^{\bar n}= \left ( \frac{1}{1+\frac{\bar n}{8} \,  \eta ^2 \cos ^2 \theta} \right )^{1/2},\\
    &&C_{-1}^{\bar n} = -i \eta \cos \theta \left ( \frac{\frac{\bar n}{8}}{1+\frac{\bar n}{8}\, \eta ^2 \cos ^2 \theta} \right )^{1/2} \nonumber ,
\end{eqnarray}
implying $|C_{-1}^{\bar n}|<<|C_{+1}^{\bar n}|$. Therefore, we can identify the positive eigenvalue correction of $V^{\bar n}$ as corresponding to the unperturbed $\ket{\vec{f}^{(0)}, \bar n , +1}$, while the negative one will correspond to the unperturbed $\ket{\vec{f}^{(0)}, \bar n , -1}$. \\
Thus, going back from the $\bar n-$notation to the $n-$notation, we can write the eigenvalue's first-order corrections
\begin{eqnarray}
 \left\{\begin{array}{ll} 
    a^{(1)}_{n,+1} =  \frac{\sin^2\theta}{\tilde E^2} \left ( 1+ \eta^2\sin^2\theta ( \frac32 +3  n +\frac{ n}{4} \cot^2\theta ) \right )\,, \\
    \;\;\;\;\;\;\;\;\;\;\;\;\;\;\;\;\;\;\;\;\;\;\;\;\;\;\;\;\;\;\;\;\;\;\;\;\;\;\;\;\;\;\;\;\;\;\;\;\;\;\;\;\;\;\;\;\;\;\;\;\;\;\;\;\;\;\;\;\;\;\;\;\;\; n>0 \\ 
    \\
    a^{(1)}_{n,-1} =  -\frac{\sin^2\theta}{\tilde E^2} \left ( 1+ \eta^2\sin^2\theta ( \frac92 +3 n +\frac{n+1}{4} \cot^2\theta ) \right )\,,\\
    \;\;\;\;\;\;\;\;\;\;\;\;\;\;\;\;\;\;\;\;\;\;\;\;\;\;\;\;\;\;\;\;\;\;\;\;\;\;\;\;\;\;\;\;\;\;\;\;\;\;\;\;\;\;\;\;\;\;\;\;\;\;\;\;\;\;\;\;\;\;\;\;\;\;  n \geq 0 
    \end{array} \right . \nonumber \\
    \label{eqa1npm}
\end{eqnarray}
thus implying, from Eq.~\eqref{eqa1npm} and Eq.~\eqref{eqa10p}, that at first order in $\lambda$ we have
\begin{eqnarray}\label{eqacompl}
    a_{n,\alpha} = 2n &&+1 -\alpha +\lambda a^{(1)}_{n,\alpha}\, ,
\end{eqnarray}
with
\begin{eqnarray}
    a^{(1)}_{n,\alpha} &=& \alpha \frac{\sin^2\theta}{\tilde E^2} \left ( 1+ 3 (n+\frac{1}{2}+ \delta_{\alpha,-1})\eta^2\sin^2\theta\right.\nonumber\\ 
    &&\left. +\frac {n+\delta_{\alpha, -1}}{4} \eta^2 \cos^2\theta \right ) \,.
    \label{eq_a_corr}
\end{eqnarray}
Here $\alpha =\pm 1$, with $\delta_{\alpha, -1}$ being the Kronecker delta.

\subsection{Perturbative corrections to the energy spectrum}
Remembering the definition Eq.~\eqref{aEdef} and the above Eq.~\eqref{eqacompl}, we can solve for the corrected energy eigenvalues, to find
\begin{eqnarray} 
E^2 _{n,\alpha}(k^3) =&&  \, m^2_f +(k^3)^2 + e B \left (2 n +1 -\alpha + \frac{M^2}{2} a^{(1)}_{n,\alpha} \right )\nonumber\\
=&& \,{{E^{(0)}_{n,\alpha}}^2(k^3)} + \frac{e B M^2}{2} a^{(1)}_{n,\alpha} \,,
\end{eqnarray}
where $m_f$ was defined in Eq.~\eqref{mf}.
Substituting Eq.\eqref{eq_a_corr}, and consistently retaining terms up to order $\eta^2$, we obtain the modified energy eigenvalues for the VSR system in the general configuration $\mathbf{B}\cdot  {\mathbf{\hat n}} = B\cos\theta$ 
\begin{eqnarray}\label{eqEna}
E^{\pm }_{n,\alpha}(k^3) &=& \pm \left[ 
{{E^{(0)}_{n,\alpha}}^2(k^3)}  + \frac{ M^2}{2} \frac{\alpha e B  \sin^2\theta}{({{E^{(0)}_{n,\alpha}}} -k^3 \cos\theta)^2}\right.\nonumber\\
&&\left.\times\left ( 1+ 3 (n+\frac{1}{2} + \delta_{\alpha,-1}) \frac{eB \sin^2\theta}{({{E^{(0)}_{n,\alpha}}} -k^3 \cos\theta)^2} \right .\right.  \nonumber\\
&& \left. \left . +\frac {n+\delta_{\alpha, -1}} 4 \frac{eB \cos^2\theta}{({{E^{(0)}_{n,\alpha}}} -k^3 \cos\theta)^2} \right )
\right]^{\frac{1}{2}} .  
\end{eqnarray}


\section{Gyromagnetic Factor and Penning Traps} \label{sec4}

In this section we discuss the above
results in the context of Penning traps. For this purpose, notation and results from~\cite{Koch:2021adn, penning1} will be used. Furthermore, while our work until now has been independent of the fermionic or leptonic family taken into account, in the following analysis we will refer to electrons.\\
Since the first experimental observation of the anomalous magnetic moment of the electron $g\neq 2$~\cite{foley,kusch}, the measurements have been continuously improved. The currently most precise direct measurement of the electron's $g$ factor has 13 significant digits~\cite{penning1,penning2,penning3}. 
These experiments are based on Penning traps.
The motion of an electron in a Penning trap has four eigenfrequencies, known as the spin-, cyclotron-, axial-, and magnetron-frequency. These four frequencies can be combined in a suitable ratio to extract an experimental value for the gyromagnetic factor of the electron, as explained in Eq.~$(10)$ from \cite{penning1}. However, for the calculations in this paper we will consider a simplified setup without electric fields, magnetron nor cavity shifts effects.
\subsection{The g-factor analysis without VSR}
For small values of the magnetic field, defining the ``free-space" cyclotron frequency $\nu_c = \frac{eB}{2 \pi m}$ \cite{penning1}, the energy-eigenvalues of the unperturbed Dirac-system are given by~\cite{brown}
\begin{eqnarray} \label{EnergiesOdom}
E^{(u)}_{n, \pm1}& \simeq &\frac{1}{2}
\left( 2 n+ 1\pm 1 \right)h \nu_c
\pm \frac{a}{2}h \nu_c -\nonumber \\
&&-\frac{1}{8}
\frac{h^2 \nu_c^2}{m c^2} \left( 2 n+ 1\pm 1 \right)^2 .
\end{eqnarray}
Here, the term $a \equiv (g-2)/2$ with $g$ being the anomalous magnetic moment, arises from adding an additional $\frac{a}{2}\sigma_{\mu \nu} F^{\mu \nu}$ term to the equations of motion, which we can think of as a perturbation arising from Quantum Field Theory (QFT) loop corrections.
Higher orders in $|\vec B|$ are experimentally not relevant~\cite{Koch:2021adn}, since
the magnetic field strength, used in \cite{penning1,penning2,penning3}, and
which can be calculated from the measured cyclotron frequency of about $\nu_c\approx 149~$GHz~\cite{penning1}, is too ``weak''
\be\label{BB}
|\vec B|=\frac{2\pi \nu_c m}{e}\approx 5.3~T \to \epsilon = \frac{eB}{m^2} \sim 10^{-9}\,,
\ee
where $e$ is the absolute value of the electron charge.\\
With our assumptions, the expression of $a$ only depends upon the anomaly frequency $\nu_a$ and the relativistic cyclotron frequency $f_c$ \cite{penning1}, which in our notation correspond respectively to the transition energies $E^{(u)}_{0,-1} -E^{(u)}_{1,+1}$ and $E^{(u)}_{1,-1}-E^{(u)}_{0,-1}$ (see also Fig.$3$ in \cite{penning1}). In fact, from \eqref{EnergiesOdom}, we directly have
\begin{equation}\label{aexp}
    \frac{E_{0,-1}^{(u)} -E_{1,+1}^{(u)}}{E_{1,-1}^{(u)}-E_{0,-1}^{(u)} + \frac{3}{2} m \epsilon^2} \rightarrow a\equiv \frac{g-2}{2} \,.
\end{equation}
Therefore, the idea is now to see what happens when considering the new VSR energy spectrum \eqref{eqEna}, which, along with the mass shift $m\to m_f$, will introduce corrections to the value of the transitions energies in the ratio \eqref{aexp}, giving
\begin{equation}\label{avsr}
    a_{VSR}\equiv \frac{g_{VSR}-2}{2} \equiv\frac{E_{0,-1} -E_{1,+1}}{E_{1,-1}-E_{0,-1} + \frac{3}{2} m_f \epsilon^2}  \,,
\end{equation}
so that, despite all the assumptions made, if VSR is correct, measuring the ratio in \eqref{avsr} would give $a_{VSR}\neq a$ already in this ideal and simplified experimental setup.

\subsection{Energy spectrum's expansion for weak magnetic field}

Let's define the new perturbative parameters $\mu = M^2/m_f^2 \ll 1$ and $\epsilon = eB/m^2_f \ll1 $. Starting from expression \eqref{eqEna} in the particle's rest frame, where we can neglect its momentum, we obtain
\begin{eqnarray}
    E_{n,\alpha} &=& m_f \bigg [  1+ \epsilon (2n +1-\alpha) \nonumber + \frac{\alpha}{2} \frac{\mu \epsilon \sin^2\theta}{1+\epsilon(2n+1-\alpha)} + \\
    && \;\;\;\;\;\;\;\;
    +\frac{3}{2} \alpha (n+\frac{1}{2} + \delta_{\alpha,-1}) \frac{\mu \epsilon^2 \sin^4 \theta}{(1+\epsilon (2n+1-\alpha))^2} + \nonumber
    \\
    && \;\;\;\;\;\;\;+\frac{\alpha (n+\delta_{\alpha, -1})}{8} \frac{\mu \epsilon^2 \sin^2 \theta \cos^2\theta}{(1+\epsilon (2n+1-\alpha))^2} \bigg ]^\frac{1}{2}  \nonumber\\
    &=& m_f ( 1+\epsilon (2n+1-\alpha))^\frac{1}{2} \times \nonumber\\
     &&\;\;\;\;\;\; \times \bigg [ 1 +\frac{\alpha}{2} \frac{\mu \epsilon \sin^2 \theta}{(1+\epsilon(2n+1-\alpha))^2} +\\ 
     &&\;\;\;\;\;\;\; +\frac{3}{2} \alpha (n+\frac{1}{2} + \delta_{\alpha,-1}) \frac{\mu \epsilon^2 \sin^4 \theta}{(1+\epsilon (2n+1-\alpha))^3} +\nonumber\\ 
     &&\;\;\;\;\;\;\;+\frac{\alpha (n+\delta_{\alpha, -1})}{8} \frac{\mu \epsilon^2 \sin^2 \theta \cos^2\theta}{(1+\epsilon (2n+1-\alpha))^3} \bigg ]^\frac{1}{2}\, ,\nonumber
\end{eqnarray}
from which we expand in the parameter $\mu$ to first order
\begin{eqnarray}
    E_{n,\alpha} &=&  m_f ( 1+\epsilon (2n+1-\alpha))^\frac{1}{2} \times \nonumber\\
     &&\;\;\;\;\;\; \times \bigg [ 1 +\frac{\alpha}{4} \frac{\mu \epsilon \sin^2 \theta}{(1+\epsilon(2n+1-\alpha))^2} \\ 
     &&\;\;\;\;\;\;\; +\frac{3}{4} \alpha (n+\frac{1}{2} + \delta_{\alpha,-1}) \frac{\mu \epsilon^2 \sin^4 \theta}{(1+\epsilon (2n+1-\alpha))^3} +\nonumber\\ 
     &&\;\;\;\;\;\;\;+\frac{\alpha (n+\delta_{\alpha, -1})}{16} \frac{\mu \epsilon^2 \sin^2 \theta \cos^2\theta}{(1+\epsilon (2n+1-\alpha))^3} \bigg ]\, ,\nonumber
\end{eqnarray}
and now in the parameter $\epsilon$ to second order
\begin{eqnarray}
    \frac{ E_{n,\alpha}}{m_f} = && \,1+\frac{\epsilon}{2} (2n+1-\alpha) -\frac{\epsilon^2}{8} (2n+1-\alpha)^2 + \nonumber\\
     &&  +\frac{\alpha}{4} \mu \epsilon \sin^2\theta \, \bigg (1-\frac{3}{2}\epsilon(2n+1-\alpha) +\\
     && +3 \epsilon (n+\frac{1}{2} + \delta_{\alpha,-1})\sin^2\theta  +\frac{\epsilon}{4} (n+\delta_{\alpha, -1}) \cos^2 \theta  \bigg ) \,. \nonumber
\end{eqnarray}
Adding usual radiative corrections from QFT as a perturbative term $\frac{a}{2} \sigma_{\mu \nu } F^{\mu\nu}$ to the VSR Dirac equation, we obtain an additional $-  \alpha  a  \epsilon \, m_f  / 2 $ term in the energy spectrum \cite{koch2022}, implying 
\begin{eqnarray} \label{appSpectrum}
    \frac{E_{n,\alpha} }{m_f} =  && 1+\frac{\epsilon}{2} (2n+1-\alpha )+ \frac{\epsilon}{4} \mu \alpha \sin^2\theta - \frac{ \epsilon}{2} \alpha a  \\
     &&  -\frac{\epsilon^2}{8} (2n+1-\alpha)^2 - \frac{3}{8}\epsilon^2 \alpha \mu  \sin^2\theta \, (2n+1-\alpha)  \nonumber\\
     &&+\frac{3}{4}\epsilon^2 \alpha \mu  (n+\frac12+\delta_{\alpha,-1}) \sin^4 \theta \nonumber\\
     && +\frac{\epsilon^2}{16} \alpha \mu (n+\delta_{\alpha, -1})  \sin^2\theta \cos^2\theta \, . \nonumber
\end{eqnarray}

\subsection{Gyromagnetic factor's corrections in VSR}
As mentioned above, the ratio of energy differences in Eq.~\eqref{aexp} is particularly useful since in the unperturbed scheme, starting from \eqref{EnergiesOdom}, it gives exactly the theoretical parameter $a$.\\
Our goal is to calculate the expression in \eqref{avsr} for our VSR approximated energy spectrum \eqref{appSpectrum}, to see if it can produce deviations from the unperturbed theoretical value $a$, derived from Lorentz invariant QFT. Keeping terms up to first order in $\mu$ and second order in $\epsilon$, we obtain 
\begin{eqnarray}
    a_{VSR} - a &=& 
    -\frac{\mu}{2} \left [ 1-3\epsilon +\frac{9}{2} \epsilon \sin^2\theta + \frac{\epsilon}{4} \cos^2\theta+ \right.\\
    && \;\;\;\;\;\;\;\; \left. +a \left (\frac{3}{2} \epsilon -\frac{3}{2}\epsilon \sin^2\theta -\frac{1}{8} \epsilon \cos^2\theta \right ) \right ] \sin^2\theta \, , \nonumber
\end{eqnarray}
from which we see that the discrepancy between the VSR value $g_{VSR}$ and the QFT's one $g$ would be
\begin{equation} 
    g_{VSR} -g = -\mu \left [ 1-
    \frac{11 }{8} ( 2 -\frac{34}{11} \sin^2\theta - a \cos^2\theta )\epsilon  \right ] \sin^2\theta  \,.
\end{equation}
Being $\epsilon$ already pretty small \eqref{BB}, in the following we will consider the relation at order $\epsilon^0$
\begin{equation} \label{gexp1}
    g_{VSR} -g \sim -\mu \sin^2 \theta \,.
\end{equation}
\subsection{Estimation of an upper limit for $\mu$}
Here, using \eqref{gexp1}, we want to put experimental bounds on the magnitude of the VSR perturbative parameter $\mu$ and, consequently, the electronic VSR mass term $M$. \\
Obviously, due to our assumptions, we cannot compare directly $g_{VSR}$ with the experimental value of the gyromagnetic factor $g_{exp}$ measured in a non-ideal Penning Trap, as done in \cite{penning1,penning2}. Nevertheless, one should not doubt that, if VSR is correct, the modification found in \eqref{gexp1} would be hidden in the value of $g_{exp}$, probably together with other terms deriving from the analysis in the complete scheme. Anyway, since our aim is just to give an upper bound for $\mu$, here we will make the further assumption that the whole discrepancy between the experimental $g_{exp}$ and theoretical value $g$ would be due to the VSR perturbation \eqref{gexp1}, i.e.
\begin{equation} \label{gexp2}
    g_{exp} -g \sim -\mu \sin^2 \theta \,.
\end{equation}
Using the most precise so far current experimental value  for  $g_{exp}/2 = 1.001159 652 180 73 (28)$ \cite{CODATA, penning1, penning2} and as the theoretical prediction $g/2=1.001 159 652 182 032 (720)$ \cite{gfactor}, we see that Eq.~\eqref{gexp2}, first of all, is consistent with the electron g-factor discrepancy since $g_{exp}-g<0$.\\
Considering again the above-cited values for $g$ and $g_{exp}$, and observing that clearly $\sin^2 \theta \leq 1$, directly gives the following restriction for the $\mu$ parameter
\begin{equation}\label{boundmu}
    \mu \lesssim 3\times 10^{-12}\, ,
\end{equation}
that is comparable or even stronger than other upper bounds found in literature \cite{dunn2006, maluf2014}. The only more stringent estimation would be $\mu < 9.7\times 10^{-19}$ obtained in~\cite{alfaro2}, which refers to electric dipole-like interaction terms $\vec n \times \vec E \cdot \vec \sigma$, where $\vec E$ represent an electric field here. Nevertheless, as already stated in \cite{dunn2006}, experiments in which the electric and magnetic fields are parallel, like the one \cite{edipole} used in \cite{alfaro2} to give the above upper bound, are insensitive to interaction terms of that type. Therefore, other experiments with no-parallel magnetic and electric field, should be used in this case to give a coherent upper limit from electric dipole-like VSR interactions.\\
At this point, by using for the electron an approximated mass of $m_f \sim 0.51\, MeV$, we can traslate the restriction in Eq.~\eqref{boundmu} to the following rough upper bound for the electron's VSR parameter
\begin{equation} \label{Melimit}
    M < \sqrt{ 10^{-13} m_f^2 } \approx 1 \, eV \, ,
\end{equation}
Observe that, the electron's VSR parameter $M$ and the electronic neutrino's VSR parameter $M_{\nu}$ in a C-symmetric VSR Dirac equation \eqref{VSRdirac} would come from the same VSR parameter included in the VSR extension of the Standard Model for each leptonic family \cite{dunn2006,alfaro2}, meaning $M_\nu=M$. Therefore, we would also have an upper limit for the VSR electronic neutrino's mass $M_\nu \lesssim 1 \, eV$, which is interestingly similar to the upper bound actually known for the electronic neutrino \cite{neutrinoup}, and leaves open the possibility for VSR to be the mechanism or one of the mechanisms giving mass to neutrinos. \\
Lastly, let's spend some words on the nature of the VSR $\theta$ angle, which we have not discussed so far. In general, we can think of two different scenarios:
\begin{itemize}
    \item The VSR spatial preferred direction $\mathbf{\hat n}$ represents some universal background effective property \cite{background}. Thus, we should take into account the orientation changes of the vector $\vec n$ respect to $\vec B$ due to Earth's motions in the Universe. For example, if the experiment's duration is of order $\sim days$, we should consider the Earth's rotation movement by averaging in $\theta\in(0,\pi)$, for which we would obtain the replacement $\sin^2\theta \to O(1) <1$.
    \item If the VSR special four-vector $n_\mu$ is a dynamic ingredient of some more fundamental theory of nature, therefore an extension of VSR including, for example, gravitational and non-inertial effects may be needed to tackle the time evolution problem of the orientation between vector $\mathbf{\hat n}$ and the magnetic field $\vec B$ accurately.
\end{itemize}
Obviously, depending on the choosen scenarios different consequencies could arise. However, since in both cases the angular effects will introduce a correction factor of $O(1)$, our rough upper limit's estimation \eqref{boundmu} for the VSR parameter would remain valid anyway. We therefore leave a more exhaustive and precise analysis of the $\theta$-dependence, like the one done in \cite{kostel}, for a future work.

\section{Conclusions} \label{concl}

In this work we have analyzed, in the framework of VSR, the effect of an homogeneous magnetic field $\vec B$ on the energy spectrum of a Dirac Fermion. First, we found the exact solution to the energy eigenvalues problem in the case of $\vec B \parallel \mathbf{\hat n}$. After that, relaxing this parallelism condition, we found the expression for the energy spectrum at first order in the VSR parameter $\mu$. By considering usual radiative QFT's corrections, we encountered an expression for the experimental parameter $a_{exp}$ in our scheme, which, confronted with the theoretical one, gave us a theory-experiment discrepancy proportional to $\mu = M^2/m_f^2$. Therefore, from the current precision in Penning trap experiments that measure the electronic $g$-factor \cite{penning1,penning2,penning3}, we found an upper bound for the VSR electron's mass $M \lesssim 1\, eV$.\\
If VSR is realized in nature as modeled in \cite{dunn2006,alfaro2}, the VSR parameter $M$ would affect leptonic doublets as a whole, generating in particular a mass $M_\nu=M$ for the electronic neutrino. The upper bound \eqref{Melimit} we found, then, is compatible with the present acceptable mass range for the electronic neutrino \cite{neutrinoup}. \\
In conclusion, we note that this work could actually also apply to the case of the muonic gyromagnetic factor, which recently is getting attention due to the increasing discrepancy between the theoretical and experimental values of its $g-$factor \cite{muon1}. However, since the experiments that measures the muonic $g-$factor use a semi-classical approach \cite{muon2}, a link between our scheme and the experimental one must still be found, to make a coherent confront. We plan to include this analysis in a future work.

\begin{acknowledgments}

A.S. acknowledges financial support from ANID Fellowship CONICYT-PFCHA/DoctoradoNacional/2020-21201387. E.M. acknowledges financial support from Fondecyt Grant No 1190361 and ANID PIA Anillo ACT192023.\\

\end{acknowledgments}

\appendix

\section{Calculation of the Equation for $\varphi(x^1)$}\label{app1}
For the sake of notation's simplicity, we define
\begin{equation} \label{defD}
    {D} = (n \cdot p )^{-1}= (E-k^3\cos\theta - p^1 \sin\theta)^{-1}  \,,
\end{equation}
to be used in this appendix. Let`s expand the second squared bracket in \eqref{eq_phi3}, keeping in mind the relation \eqref{anticom2} and the fact that mixed $\sigma$'s products which do not involve operators simply get cancelled due to their anticommutation relations 
\begin{widetext}
\begin{eqnarray} \label{app1.1}
&&\left[\left( E - \frac{M^2}{2} D \right)^2 - m^2 -(p^1 - \frac{M^2 \sin\theta}{2} D)^2-(k^2 - e B x^1)^2- (k^3 - \frac{M^2 \cos\theta}{2}D)^2 \right.\\
&& \left. - i e B  \sigma^1  \sigma^2 - e B \frac{M^2}{2} \sin\theta D x^1  \sigma^1  \sigma ^2 - e B \frac{M^2}{2} \sin\theta x^1 D   \sigma^2  \sigma ^1  - e B \frac{M^2}{2} \cos\theta D x^1  \sigma^3  \sigma^2 -e B \frac{M^2}{2} \cos\theta  x^1 D  \sigma^2  \sigma^3\right]\varphi(x^1)  = 0 \, . \nonumber
\end{eqnarray}
\end{widetext}

Therefore, we now have to calculate $D (x^1)$. To do that, we will use the integral representation \eqref{intrepr}, and we define for simplicity $\tilde E = E -k^3 \cos\theta$, implying
\begin{eqnarray}
      D (x^1 \varphi) &=& \int_0^\infty dt \; exp[-t (\tilde E - p^1 \sin\theta)] \, x^1 \varphi \\
      &=& \int_0^\infty dt \; e^{-t \tilde E} \sum_{j=0}^\infty (t \sin\theta  p^1)^j x^1 \varphi \nonumber \,,
\end{eqnarray}   
which can be expanded as
\begin{eqnarray}
   D (x^1 \varphi)   &=& x^1 \int_0^\infty dt \; e^{-t \tilde E} \sum_{j=0}^\infty (t \sin\theta  p^1)^j \, \varphi\\
      && -i \sin\theta \int_0^\infty t dt \; e^{-t \tilde E} \sum_{j=1}^\infty (t \sin\theta p^1)^{j-1} \varphi \nonumber \, .
\end{eqnarray}
Sending the index of the second sum to $i \rightarrow i-1$, and ``reversing'' the use of the integral representation, we have the relation
\bea
     D(x^1 \varphi ) &=& x^1 D \varphi +i\sin\theta \frac{d}{d\tilde E}(D) \varphi \,,
\eea
and considering the definition \eqref{defD}
\begin{eqnarray}
    D(x^1 \varphi ) &=& x^1 D \varphi -i\sin\theta D ^2 \varphi \, .
\end{eqnarray}
Therefore, since the terms $x^1 D \varphi(x^1)$ always get cancelled from the other terms with inverted $\sigma$ product, and since
\begin{equation}
    \sigma^i  \sigma^j = i \epsilon_{ijk}  \sigma^k \, ,
\end{equation}
Eq. \eqref{app1.1} then becomes
\begin{widetext}
\begin{eqnarray}\label{Aeq1}
    &&\left[\left( E - \frac{M^2}{2} D \right)^2 - m^2 -(p^1 - \frac{M^2 \sin\theta}{2} D)^2-(k^2 - e B x^1)^2- (k^3 - \frac{M^2 \cos\theta}{2}D)^2 \right.\\
    && \left. + e B  \sigma^3 - e B \frac{M^2}{2} \sin^2\theta D^2   \sigma^3  +e B \frac{M^2}{2} \sin\theta \cos\theta D^2  \sigma^1 \right]\varphi(x^1)  = 0 \, . \nonumber
\end{eqnarray}
\end{widetext}
Finally, expanding the squares involving the terms $E$, $p^1$ and $k^3$, and remembering that $N_\mu  p^\mu = D \, n_\mu  p^\mu = 1$ while $n_\mu n^\mu = 0$, we get to Eq. \eqref{eq_phi4}, that we have already seen in the main text of this article.\\
Note that Eq.~\eqref{Aeq1} has the correct limits: in fact, sending $B\to 0$, we re-obtain, as expected, the usual VSR dispersion relation
\begin{equation}
    [E^2 - p^2 -m^2 -M^2 ]\, \varphi = 0\,,
\end{equation}
while, for $\theta \to 0$ we find again Eq.~\eqref{eq_phi2}, the equation of motion for the case $\vec B \parallel \mathbf{\hat n}$.


\section{Calculation of the integrals $I_1( \bar n,k)$ and $I_2(\bar n,k)$} \label{app2}

In the main text, we defined the integrals
\begin{eqnarray}
I_1 (\bar n,k) = \int_{-\infty}^{\infty}d\xi\,e^{-\xi^2/2}H_{\bar n}(\xi)\partial_{\xi}^k\left( e^{-\xi^2/2} H_{\bar n}(\xi)  \right)\nonumber ,\\
\label{eq_In}
\end{eqnarray}
where $H_{\bar n}(z)$ are the Hermite polynomials, defined by
\begin{eqnarray}
H_{\bar n}(z) = (-1)^{\bar n}
e^{z^2}\partial_{z}^{\bar n}\left( e^{-z^2} \right)
\end{eqnarray}
In particular, we remark that $H_{0}(z) = 1$, and hence for the particular case $\bar n=0$, 
Eq.~\eqref{eq_In} reduces to
\begin{eqnarray}
I_1(0,k) &=& \int_{-\infty}^{\infty}d\xi\,e^{-\xi^2/2}\partial_{\xi}^{k}\left( e^{-\xi^2/2} \right)\nonumber\\
&=& 2^{-k/2}(-1)^k
\int_{-\infty}^{\infty}d\xi\,e^{-\xi^2}\,H_{k}(\xi/\sqrt{2})
\end{eqnarray}
From the standard result (see for instance Gradshteyn-Rhizik \cite{grad}, p. 837)
\begin{eqnarray}
&&\int_{-\infty}^{\infty}e^{-(x-y)^2}H_{k}(\alpha x)dx = \\
&& \;\;\; = \sqrt{\pi}(1 - \alpha^2)^{k/2}H_{k}\left(\frac{\alpha y}{\sqrt{1 - \alpha^2}}  \right),\nonumber\\
\end{eqnarray}
we obtain (setting $\alpha = 1/\sqrt{2}$ and $y=0$)
\begin{eqnarray}
I_1(0,k)= \sqrt{\pi}(-1)^k
2^{-k}H_{k}(0).
\end{eqnarray}
Given that $H_{2n+1}(0)= 0$, while $H_{2n}(0) = (-1)^n \Gamma(2n+1)/\Gamma(n+1)$, we obtain 
\begin{eqnarray}
&&I_1(0,2k) = \sqrt{\pi}2^{-2k}(-1)^k\frac{\Gamma(2k+1)}{\Gamma(k+1)},\\
&&I_1(0,2k+1) = 0 \,\nonumber .
\end{eqnarray}
Let's now focus on the case $\bar n>0$. For this purpose, we notice that the functions
\begin{eqnarray}
\langle\xi|\varphi_{\bar n}\rangle = \varphi_{\bar n}(\xi) = \frac{1}{\sqrt{2^{\bar n} \bar n! \sqrt{\pi}}}e^{-\xi^2/2}H_{\bar n}(\xi)
\end{eqnarray}
constitute a complete orthonormal set (they are just the eigenfunctions of the 1D-harmonic oscilator)
\begin{eqnarray}
&&\sum_{\bar n=0}^{\infty}|\varphi_{\bar n}\rangle\langle\varphi_{\bar n}| = \mathbf{1} \\
&&\langle\varphi_{\bar n}|\varphi_{\bar n'}\rangle = \int_{-\infty}^{\infty}d\xi \langle\varphi_{\bar n}|\xi\rangle\langle\xi|\varphi_{\bar n'}\rangle = \delta_{\bar n,\bar n'}\nonumber
\end{eqnarray}
With these definitions, along with the momentum operator $p_{\xi}= -i\partial_{\xi}$, we have the rather simple expression
\begin{eqnarray}
I_1(\bar n,k) = 2^{\bar n} \bar n! \sqrt{\pi} (i)^k \langle \varphi_{\bar n}|(p_{\xi})^k|\varphi_{\bar n} \rangle
\label{eq_In1}
\end{eqnarray}
Since the momentum (Fourier) eigenbasis is also complete, we have
\begin{eqnarray}
&&\int_{-\infty}^{\infty}dp|p\rangle\langle p| = \mathbf{1} \, ,\,\,\,\langle\xi | p\rangle = \frac{1}{\sqrt{2\pi}}e^{i\xi p}.
\end{eqnarray}
Therefore, we can insert the identity in the basis of momentum eigenstates into Eq.~\eqref{eq_In1} to obtain
\begin{eqnarray}
I_1(\bar n,k) &=& 2^{\bar n} \bar n! \sqrt{\pi} (i)^k \int_{-\infty}^{\infty}dp\langle\varphi_{\bar n}|(p_{\xi})^k|p\rangle\langle p|\varphi_{\bar n} \rangle\nonumber\\
&=& 2^{\bar n} \bar n! \sqrt{\pi} (i)^k \int_{-\infty}^{\infty}dp\,p^{k}\left|\langle p|\varphi_{\bar n} \rangle\right|^2
\label{eq_In2}
\end{eqnarray}
Finally, notice that the functions $\langle p|\varphi_{\bar n}\rangle$ are related to the functions $\langle\xi|\varphi\rangle$ via Fourier representation, since
\begin{eqnarray}
\langle p|\varphi_{\bar n}\rangle &=& \int_{-\infty}^{\infty}d\xi \langle p|\xi\rangle\langle\xi|\varphi_{\bar n}\rangle\nonumber\\
&=& \frac{1}{\sqrt{2\pi}}\int_{-\infty}^{\infty} d\xi \,e^{-i \xi p}\varphi_{\bar n}(\xi)\nonumber\\
&=& \frac{1}{\sqrt{2^{\bar n} \bar n! \sqrt{\pi}}} \frac{1}{\sqrt{2\pi}}\int_{-\infty}^{\infty}d\xi \,e^{-i \xi p}e^{-\xi^2/2}H_{\bar n} (\xi)\nonumber\\
&=& \frac{(-i)^{\bar n}}{\sqrt{2^{\bar n} \bar n! \sqrt{\pi}}}e^{-p^2/2}H_{\bar n} (p)
\end{eqnarray}
Substituting this last result into Eq.~\eqref{eq_In2}, we have
\begin{eqnarray}
I_1(\bar n,2k) = (-1)^k \int_{-\infty}^{\infty}dp\, p^{2k} e^{-p^2} \left (H_{\bar n} (p)\right )^2,
\label{eq_In3}
\end{eqnarray}
with $I_1(n,2k+1) = 0$ trivially by parity.\\
Using the Hermite-Fourier series representation for the power $p^{2k}$, we have
\begin{eqnarray}
p^{2k} = \frac{(2k)!}{2^{2k}}\sum_{\ell = 0}^{k}\frac{H_{2\ell}(p)}{(2\ell)!(k-\ell)!}, 
\end{eqnarray}
and inserting this expression into Eq.~\eqref{eq_In3} we obtain
\begin{eqnarray}
I_1(\bar n,2k) &=& (-1)^k\frac{(2k)!}{2^{2k}}
\sum_{\ell=0}^k \frac{1}{(2\ell)!(k-\ell)!}\\
&&\times\int_{-\infty}^{\infty}dp\,e^{-p^2} \left(H_{\bar n} (p)\right)^2H_{2\ell}(p) \nonumber
\label{eq_In4}
\end{eqnarray}
We can evaluate this last integral by applying the identity (see Gradshteyn-Rhizik \cite{grad}, p.797, for $s = (\bar n + m + k)/2$)
\begin{eqnarray} \label{rhizikHHH}
\int_{-\infty}^{\infty}dz\,e^{-z^2} H_{k}(z)H_{m}(z)H_{\bar n}(z) = \frac{2^{s}\sqrt{\pi} \, k! \, m! \, \bar n!}{(s-k)!(s-m)!(s-\bar n)!}\nonumber\\
\end{eqnarray}
Applying this result into Eq.~\eqref{eq_In4}, we obtain
\begin{eqnarray}
I_1(\bar n,2k) &=& \frac{\sqrt{\pi}(-1)^k (2k)! \, \bar n! \, 2^{\bar n-2k}}{k!}\sum_{\ell=0}^{k}2^{\ell}\left(\begin{array}{c} \bar n \\\ell \end{array}\right)\left(\begin{array}{c}k\\\ell \end{array}\right)\nonumber\\
&=&  \sqrt{\pi} (-1)^k \bar n! \, 2^{\bar n-2k}\frac{\Gamma(2k+1)}{\Gamma(k+1)} F(-k,-\bar n;1;2)\nonumber\\
\end{eqnarray}
where $F(a,b;c;z)$ is the Hypergeometric function.\\
Furthermore, we want to calculate the integrals of the form
\begin{equation}\label{i2def}
    I_2 (\bar n,k) =\int_{-\infty}^{\infty} d\xi e^{-\xi^2/2} H_{\bar n}(\xi) \partial^k_\xi \left ( e^{-\xi^2/2} H_{\bar n-1}(\xi) \right ) \,.
\end{equation}
We can undergo all the previous steps done for the integrals $I_1$ to find
\begin{eqnarray}
    I_2 (\bar n, k) &=& 2^{\bar n-\frac12}  (\bar n-1)!  \sqrt{\pi n} (i)^{k} \langle \varphi_{\bar n}|(p_{\xi})^k|\varphi_{\bar n-1} \rangle = \nonumber \\
    &=& i^{k+1} \int_{-\infty}^{\infty} dp \, p^k e^{-p^2} H_{\bar n}(p) H_{\bar n -1}(p)
    \,,
\end{eqnarray}
which, due to the parity of the integrand is easily seen to be 0 for even values of $k$: $I_2 (\bar n,2k) = 0$. Remembering that for odd powers of momentum $p$
\begin{equation}
    p^{2k+1} = \frac{(2k+1)!}{2^{2k+1}} \sum_{\ell = 0}^{k}\frac{H_{2\ell+1}(p)}{(2\ell+1)!(k-\ell)!}, 
\end{equation}
we find
\begin{eqnarray}
    &&I_2 (\bar n, 2k+1) = (-1)^{k+1} \frac{(2k+1)!}{2^{2k+1}} \times \\
    && \;\;\;\;\;\; \times \sum_{\ell=0}^k \frac{1}{(2\ell+1)!(k-\ell)!} \int_{-\infty}^{\infty} dp \, e^{-p^2} H_{2l+1} H_{\bar n} H_{\bar n -1}\,. \nonumber 
\end{eqnarray}
Using again the expression \eqref{rhizikHHH} for the integral in $I_2$, we finally obtain
\begin{eqnarray}\label{i2expr}
    I_2 (\bar n, 2k+1) &=& \sqrt{\pi} (-1)^{k+1}  \frac{ (2k+1)!}{k!} (\bar n -1) ! \, 2^{\bar n -2k-1} \times \nonumber\\
    && \times \sum_{\ell=0}^k 2^{\ell}\left(\begin{array}{c}  k  \\\ell \end{array}\right)\left(\begin{array}{c}\bar n \\\ell +1\end{array}\right) =\\
    &=&\sqrt{\pi} (-1)^{k+1}  (\bar n -1) ! \, 2^{\bar n -2k-1} \frac{ \Gamma(2k+2)}{\Gamma(k+1)}  \times \nonumber\\
    && \times \frac{F(-k-1;-n;1;2)-F(-k;-n;1;2)}{2} \nonumber\, .
\end{eqnarray}


\section{Calculation of Matrix elements $V^{\bar n}_{\alpha,\alpha'}$}\label{app3}

First of all, we observe that the perturbation $   V $ is hermitian and therefore also $V^{\bar n}$ is hermitian, implying $V^{\bar n}_{+1,-1}=(V^{\bar n}_{-1,+1})^*$. Furthermore, since the diagonal elements in $   V$ are equal but opposite in sign, we will have $V^{\bar n}_{+1,+1}=-V^{\bar n}_{-1,-1}$. Therefore, is sufficient to calculate only two matrix elements: $V^{\bar n}_{+1,+1}$ and $V^{\bar n}_{+1,-1}$.\\

\subsection{Case $V^{\bar n}_{+1,+1}$}

We want to calculate the matrix element
\begin{eqnarray}
    V^{\bar n}_{+1,+1} &=& \bra{\vec f^{(0)} ,\bar n ,+1}{V}\ket{ \vec f^{(0)} , \bar n ,+1} \, .
\end{eqnarray}
Inserting two identities through the completeness relation $1= \int d\xi \ket{\xi}\bra{\xi}$ we get
\begin{eqnarray}\label{a1.1}
V^{\bar n}_{+1,+1}  = &&  \frac{\sin^2\theta}{2^{\bar n}\bar  n! \sqrt{\pi} } \int_{-\infty}^{\infty} d\xi \; e^{-\xi^2/2} H_{\bar n}(\xi) \frac{1}{ P^2_\xi} (e^{-\xi^2/2} H_{\bar n}(\xi)) = \nonumber\\
&& = - \frac{\sin^2\theta}{2^{\bar n} \bar n! \sqrt{\pi} \tilde E^2 } \frac{d}{d A }\int_{-\infty}^{\infty} d\xi \; e^{-\xi^2/2} H_{\bar n}(\xi) \times \nonumber\\
&& \;\;\;\times \frac{1}{A+i \eta\sin\theta \partial_\xi} (e^{-\xi^2/2} H_{\bar n}(\xi)) \bigg |_{A=1} \, ,
\end{eqnarray}
where we have defined the dimensionless quantity $\eta={\sqrt{e B} }/{\tilde E}$. By representing the inverse operator in \eqref{a1.1} with the integral form, we obtain
\begin{eqnarray}\label{a1.2}
V^{\bar n}_{+1,+1}  = && - \frac{\sin^2\theta}{2^{\bar n} \bar n! \sqrt{\pi} \tilde E^2 } \frac{d}{d A}\int_{0}^\infty dt \int_{-\infty}^{\infty} d\xi \; e^{-\xi^2/2} H_{\bar n}(\xi) \nonumber \\
&& \;\;\; \times  e^{- A t (1+i \eta\sin\theta \partial_\xi)} (e^{-\xi^2/2} H_n(\xi)) \bigg |_{A=1} =  \nonumber\\
=&& - \frac{ \sin^2\theta}{2^{\bar n} \bar n! \sqrt{\pi} \tilde E^2 } \frac{d}{dA}\int_{0}^\infty e^{-At} dt \int_{-\infty}^{\infty} d\xi \; e^{-\xi^2/2} H_{\bar n}(\xi) \nonumber \\
&& \;\;\; \times  e^{-i t \eta\sin\theta  \partial_\xi} (e^{-\xi^2/2} H_{\bar n}(\xi)) \bigg |_{A=1} \,,
\end{eqnarray}
and expanding the exponential operator
\begin{eqnarray}\label{a1.3}
V^{\bar n}_{+1,+1} =&& - \frac{\sin^2\theta}{2^{\bar n} \bar n! \sqrt{\pi} \tilde E^2 }  \sum_{k=0}^\infty \frac{(-i \eta\sin\theta )^k}{k!} \frac{d}{dA} \int_{0}^\infty e^{-At} t^k dt \nonumber  \\
&& \times  \int_{-\infty}^{\infty} d\xi \; e^{-\xi^2/2} H_{\bar n}(\xi) \partial_\xi^k (e^{-\xi^2/2} H_{\bar n}(\xi)) \bigg |_{A=1} = \nonumber\\
=&&  \frac{ \sin^2\theta}{2^{\bar n} \bar n! \sqrt{\pi} \tilde E^2 }  \sum_{k=0}^\infty (k+1) (i \eta\sin\theta )^k  I_1(\bar n,k) \, ,
\end{eqnarray}
where we introduced the integral $I_1(\bar n,k)$ defined in \eqref{I1nk}, which vanish for odd values of k. Thus, we can re-write equation \eqref{a1.3} as
\begin{equation} \label{a1sumk}
    V^{\bar n}_{+1,+1}  = \frac{ \sin^2\theta}{2^{\bar n} \bar n! \sqrt{\pi} \tilde E^2 }  \sum_{k=0}^\infty (2k+1) (-1)^{k}\left(\eta\sin\theta\right)^{2k}  I_1(\bar n,2k) \, ,
\end{equation}
that is a completely real expression as expected.\\
Therefore the final expression for the $V^{(\bar n)}_{+1,+1}$ matrix element will be
\begin{equation}
    V^{\bar n}_{+1,+1}  = \frac{\sin^2\theta}{ \tilde E^2 }  \sum_{k=0}^\infty \left (\frac{\eta\sin\theta}{2} \right )^{2k} \frac{\Gamma(2k+2)}{\Gamma(k+1)} F(-k,-\bar n;1;2)\, ,
\end{equation}
which for weak magnetic field can be expanded as
\begin{equation}
    V^{\bar n}_{+1,+1}  = \frac{\sin^2\theta}{ \tilde E^2 } \left ( 1 + (2\bar n +1) \, \frac{3}{2}\left(\eta\sin\theta\right)^2 + O(\eta^4)\right )\, .
\end{equation}

\subsection{Case $V^{\bar n}_{+1,-1}$}

We want to calculate the matrix element
\begin{eqnarray}
    V^{\bar n}_{+1,-1} &=& \bra{\vec f^{(0)} ,\bar n ,+1}{V}\ket{ \vec f^{(0)} , \bar n ,-1} \\
    &=& \frac{\sin \theta \cos \theta}{\sqrt{\pi} 2^{\bar n-\frac12} (\bar n-1)! \sqrt{\bar n}}  \int_{-\infty}^{\infty} d\xi \; e^{-\xi^2/2} H_{\bar n}(\xi) \times \nonumber\\
    && \times \frac{1}{ P^2_\xi} (e^{-\xi^2/2} H_{\bar n -1}(\xi))\nonumber \,.
\end{eqnarray}
Following exactly the same procedure as before, we get to the expression
\begin{eqnarray}
    V^{\bar n}_{+1,-1} &=& \frac{\sin \theta \cos \theta}{\sqrt{\pi} 2^{\bar n-\frac12} (\bar n-1)! \sqrt{\bar n}} \frac{1}{\tilde E^2}\nonumber\\
    &&\times\sum_k (k+1)(-i \eta\sin\theta)^k I_2(\bar n ,k)\,,
\end{eqnarray}
where we introduced the integrals $I_2$ defined in \eqref{i2def}. \\
In the end, using Eq.~\eqref{i2expr} for $I_2$ and remembering that it is non-zero only for odd values of $k$, we have
\begin{eqnarray} 
   V^{\bar n}_{+1,-1} &=& i \sqrt{\frac{1}{8 \bar n}} \frac{\sin \theta \cos \theta}{\tilde E^2} \sum_{k=0}^{\infty} \left ( \frac{\eta\sin\theta}{2} \right )^{2k+1}  \frac{\Gamma(2k+3)}{\Gamma(k+1)} \times \nonumber\\
    && \times \left [ F(-k-1;-n;1;2)-F(-k;-n;1;2) \right]  \,.  \nonumber\\
\end{eqnarray}
For our purposes, when using the weak field approximation $\eta\ll1$, it will be sufficient to consider only the first term of the $k-$series, since the next one would already be order $\propto \eta^3$, obtaining
\begin{eqnarray} 
   V^{\bar n}_{+1,-1} &\approx& i  \eta \sqrt{\frac{\bar n}{2}} \frac{\sin^2 \theta \cos \theta}{\tilde E^2} \,.
\end{eqnarray}

\section{Borel regularization}\label{Borel}
In this section, we show in detail the procedure to obtain a Borel regularization for the infinite series Eq.~\eqref{a1n0sumk} defined in the main text
\begin{eqnarray} 
    a^{(1)} _{\bar 0,+1} &=& \frac{ \sin^2\theta}{ \sqrt{\pi} \tilde E^2 }  \sum_{k=0}^\infty (2k+1)(-1)^k \left(\eta\sin\theta \right)^{2k}  I_1(0,2k)\nonumber\\
    &=& \frac{ \sin^2\theta}{ \tilde E^2 } \sum_{k=0}^\infty \left(\frac{\eta\sin\theta}{2}\right)^{2k}\frac{\Gamma(2k + 2)}{\Gamma(k+1)} \, .
\label{eq_a11}
\end{eqnarray}
First, notice that for $k$ a positive integer, the ratio
\begin{eqnarray}
\frac{\Gamma(2k+2)}{\Gamma(k+1)} = \frac{(2k + 1)!}{k!} \,,
\end{eqnarray}
and let us consider, with a simplified notation $z = \left(\eta\sin\theta/2\right)^2$, the equivalent series
\begin{eqnarray}
A(z) = \sum_{k=0}^{\infty}z^k \frac{(2k + 1)!}{k!}\, ,
\label{Aseries}
\end{eqnarray}
where clearly Eq.\eqref{eq_a11} corresponds to
\begin{eqnarray}
a^{(1)} _{\bar 0,+1} = \frac{\sin^2\theta}{\tilde{E}^2}A(\left(\eta\sin\theta/2\right)^2)\,.
\end{eqnarray}
The definition of the Borel transform of this series leads to the expression
\begin{eqnarray}
BA(zt) &=& \sum_{k=0}^{\infty}\frac{(z t)^k}{k!} \frac{(2k + 1)!}{k!}\nonumber\\
&=& \frac{1}{\left(1 - 4 z t \right)^{3/2}}\,,
\end{eqnarray}
where the second result follows as an identity directly from the Taylor expansion.
As the following step, we recover the regularized expression for the series $\tilde{A}(z) \simeq A(z)$ by performing the integral transform
\begin{eqnarray}
\tilde{A}(z) &=& \int_0^{\infty}e^{-t}
BA(zt) \, dt\nonumber\\ 
&=& \int_0^{\infty}\frac{e^{-t}}{\left(1 - 4 z t \right)^{3/2}} \, dt\,.
\end{eqnarray}
In order to evaluate this integral, we perform the change of variables
$u = t - \frac{1}{4z}$, that implies $-1/(4z) \le u < \infty$, and $du = dt$. Therefore, we have
\begin{eqnarray}
\tilde{A}(z) &=& \frac{e^{\frac{-1}{4z}}}{(-4z)^{3/2}}\int_{-\frac{1}{4z}}^{\infty}u^{-3/2} e^{-u} du\nonumber\\
&=& \frac{e^{\frac{-1}{4z}}}{(-4z)^{3/2}}\Gamma\left(-\frac{1}{2},-\frac{1}{4 z} \right),
\end{eqnarray}
where in the final result we applied the definition of the Incomplete Gamma function
\begin{eqnarray}
\Gamma(s,x) = \int_{x}^{\infty}u^{s-1}e^{-u} du \,.
\end{eqnarray}
\nocite{*}

\bibliography{apssamp}

\end{document}